\documentclass[10pt,twocolumn,letterpaper]{article}

\usepackage[pagenumbers]{cvpr} 

\usepackage{graphicx}
\usepackage{amsmath}
\usepackage{amssymb}
\usepackage{booktabs}
\usepackage{enumitem}

\usepackage[pagebackref,breaklinks,colorlinks]{hyperref}

\usepackage[capitalize]{cleveref}
\crefname{section}{Sec.}{Secs.}
\Crefname{section}{Section}{Sections}
\Crefname{table}{Table}{Tables}
\crefname{table}{Tab.}{Tabs.}

\def\cvprPaperID{*****} 
\def\confName{CVPR}
\def\confYear{2022}

\begin{document}

\title{RestoreX-AI: A Contrastive Approach towards Guiding Image 
Restoration via Explainable AI Systems}

\author{Aboli Marathe\\
Symbiosis Centre for Applied \\ Artificial Intelligence (SCAAI),\\
Symbiosis International Deemed\\
University (SIU), India\\
SCTR’s Pune Institute of \\
Computer Technology, India \\
{\tt\small aboli.rajan.marathe@gmail.com}
\and
Pushkar Jain\\
Symbiosis Centre for Applied \\ Artificial Intelligence (SCAAI),\\
Symbiosis International Deemed\\
University (SIU), India\\
SCTR’s Pune Institute of \\
Computer Technology, India \\
{\tt\small pushrjain@gmail.com}
\\
\and
Rahee Walambe\\
Symbiosis Centre for Applied \\
Artificial Intelligence (SCAAI), \\
Symbiosis International Deemed \\
University (SIU), India \\
Symbiosis Institute of Technology, \\
SIU, India \\
\\
{\tt\small rahee.walambe@sit.edu.in}
\and
Ketan Kotecha\\
Symbiosis Centre for Applied \\
Artificial Intelligence (SCAAI), \\
Symbiosis International Deemed \\
University (SIU), India \\
Symbiosis Institute of Technology, \\
SIU, India \\
\\
{\tt\small director@sit.edu.in}
}
\maketitle


\begin{abstract}
Modern applications such as self-driving cars and drones rely heavily upon robust object detection techniques. However, weather corruptions can hinder the object detectability and pose a serious threat to their navigation and reliability. Thus, there is a need for efficient denoising, deraining, and restoration techniques. Generative adversarial networks and transformers have been widely adopted for image restoration. However, the training of these methods is often unstable and time-consuming. Furthermore, when used for object detection (OD), the output images generated by these methods may provide unsatisfactory results despite image clarity. In this work, we propose a contrastive approach towards mitigating this problem, by evaluating images generated by restoration models during and post training.  This approach leverages OD scores combined with attention maps for predicting the usefulness of restored images for the OD task. We conduct experiments using two novel use-cases of conditional GANs and two transformer methods that probe the robustness of the proposed approach on multi-weather corruptions in the OD task. Our approach achieves an averaged 178  increase in mAP between the input and restored images under adverse weather conditions like dust tornadoes and snowfall. We report unique cases where greater denoising does not improve OD performance and conversely where noisy generated images demonstrate good results. We conclude the need for explainability frameworks to bridge the gap between human and machine perception, especially in the context of robust object detection for autonomous vehicles.

\end{abstract}

\section{Introduction}
\label{sec:intro}

Sensors are employed by unmanned autonomous vehicles to navigate through their surroundings,  with a substantial dependence on vision-based sensors like RGB cameras. As these sensors are impacted by bad weather conditions, perception pipelines require considerable training on diverse data to increase the robustness on downstream tasks. One specific scenario which causes distortion of images is adverse weather conditions, like heavy snowfall, haze and dust tornados. In these critical situations, weather corruptions can hinder the object detectability and pose a serious threat to navigation and reliability. Thus, there is a need for efficient denoising, deraining and restoration techniques.

However, denoising techniques are often evaluated using Image similarity metrics like PSNR, SSIM \cite{a37} and not by their effectiveness in achieving results for the targeted application. It is possible that the output image of these methods has high image quality but is contextually irrelevant for the object detection task. In this work we evaluate the effectiveness of restoration techniques for denoising images with the intention of better object detection. By introducing a contrastive approach towards restoration evaluation,  a method for guiding the training of GANs and restoration progress is proposed. Additionally,  attention maps are leveraged for understanding  why these techniques assist object detection and why certain classes are easily recognized or ignored.

Primary contributions of this work are :
\begin{enumerate}
    \item Exploring two new data-based generative adversarial network techniques for denoising weather corrupted images. 
    \item Proposing a novel contrastive approach using a weighted loss for evaluating the training progress and post-training performance of the highlighted restoration techniques. 
    \item Explaining the training progress of the proposed generative denoising methods using attention maps and validating the results using object detection evaluation metrics.
    \item Evaluating the optimal noise level of the trained Restormer denoising methods (color and grayscale) using attention maps and validating the results using object detection evaluation metrics.
\end{enumerate}

\begin{figure*}
\begin{center}
\includegraphics[width=\textwidth,height=5in]{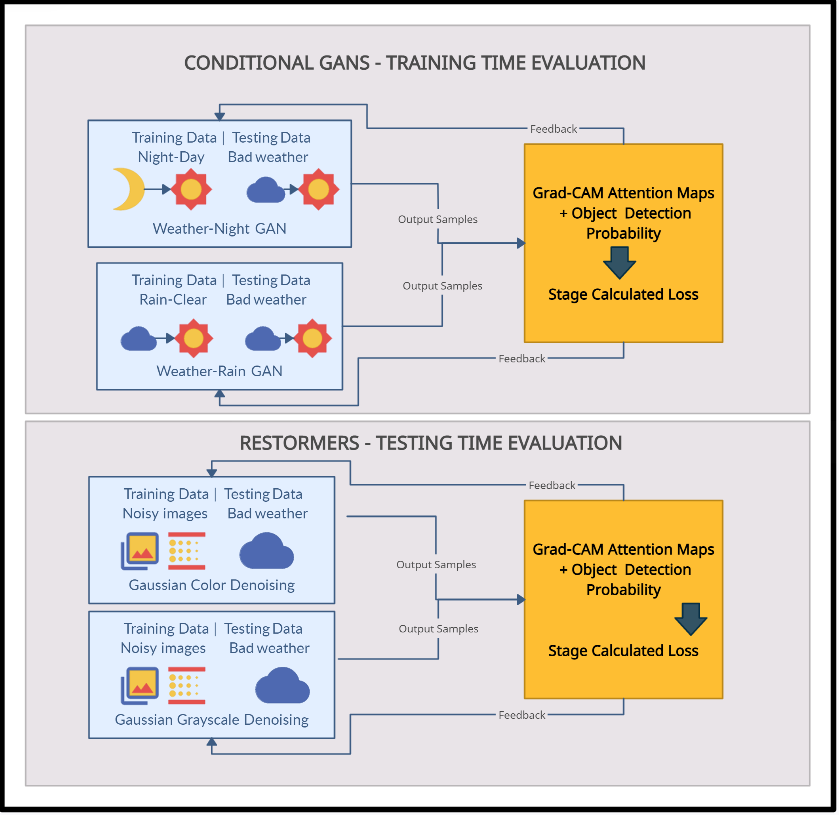}
\end{center}
  \caption{RestoreX-AI: Proposed contrastive approach for evaluating restoration models.}
\label{fig1}
\end{figure*}

\section{Related Work}

The rapid development and deployment of autonomous vehicles have exposed several critical challenges in computer vision, one of which being object detection robust to weather corruptions. At the foundation of this challenge is the classic vision task of object detection, the progress of which we summarize in the following sub-section. Ultimately, when these modules are integrated in autonomous driving systems, their suggestively “black-box” nature, triggers the concerns of industry professionals, car-makers and users alike. Without human-tailored explanations for the vehicle’s behavior, even the most advanced systems fail to benefit from widespread adoption due to concerns about safety and reliability. Especially in adverse weather, when even human drivers exercise extra caution, these robust models need to provide irrefutable explanations for the detections made at every state. We discuss the progress of previous works in these 3 directions which set the background of our work in restoration, detection and explainability.

\subsection{Multi-weather corruption and restoration}

The thermal variations accompanying weather change can adversely impact the optical, electronic, and mechanical components used in capturing visual data, thus harming the performance of visual recognition systems \cite{a4}. Frigid temperatures, snowfall or dense fog, for example, can cause condensation on the lens, further blurring the view and obscuring the object boundaries; rain streaks on car windows can generate glares or act as a double lens \cite{a36}. For an autonomous car, it is critical and essential to overcome the effects of weather conditions to ensure reliability. Number of approaches for this have been reported. For example, one study \cite{a5} looked at the performance of gated cameras, while another \cite{a6} expanded the research to include stereo, gated, and thermal cameras, as well as Radar and LiDAR scanners, and found considerable increases in car recognition in varying levels of fog, and other adverse weather conditions. The use of unique methodologies like domain adaptation to transform the weather conditions while keeping objects of interest intact. For example, \cite{a7} investigates the consequences of synthetic weather images on road segmentation and traffic object detection, whereas \cite{a8} shows that using synthetic time-of-day (night imagery) improves localization, and \cite{a9} proposes a de-raining model to improve semantic segmentation. For their efficiency in image restoration and effectiveness in removing weather corruptions, image restoration employing restomers \cite{a10} and image denoising algorithms \cite{a11} are becoming increasingly popular.

\subsection{Object Detection}

The introduction of fast detectors like SSD, Faster RCNN, and YOLO transformed the face of object detection \cite{a22,a23,a24}. Taking 2D detectors forward, 3D detection expanded with Stereo RCNN \cite{a25}, AVOD \cite{a26}, MVLidarNet \cite{a27} and MVF algorithm \cite{a28} bringing new perspectives to the task of object detection. Specific tasks like multiscale object detection \cite{a29}, pedestrian detection in crowds \cite{a30} and detection under adverse weather \cite{a31} have also been solved using ensemble methods and data augmentation.

\subsection{Explainability in Object Detection}

The early rise of deployable machine learning technologies was accompanied by criticism of the “black-box”-like nature of ML models. Particularly in sensitive applications like medical diagnosis, self-driving cars and algorithmic test checking, there is a need for thorough explainability in models to ensure the trust and safety of users. To meet these requirements, many techniques were proposed for explaining models \cite{a12,a13,a14,a15,a16}. In \cite{a17}, explainability is demonstrated by studying what each of its neurons has learned to detect. \cite{a18} focused on individual predictions, using the technique of heatmaps to highlight important pixels. Some works also interpret classifiers by identifying representative training examples \cite{a19,a20}. \cite{a21} introduced a new perspective to this challenge by making CNN-based models more transparent by producing visual explanations. As newer machine learning systems are rapidly adopted, the demand for explainable models which incorporate diverse approaches is gaining attention of the community.

\section{Methodology}

This work proposes a contrastive approach (loss) for monitoring the training progress of restoration models in the context of object detection. To put forward a diverse range of training samples, new restoration techniques using GANs have also been proposed. First, the restoration models were trained on different tasks and tested on the DAWN dataset \cite{a35}. As the models trained, their progress was monitored using the contrastive approach and simultaneously attention maps were generations to support the detection task. Finally, the OD performance of all the methods is compared using standard evaluation metrics (class AP and mAP). The 4 restoration techniques experimented with in this study are:

\begin{enumerate}
    \item Weather-NightGAN:Conditional GANs trained on night-to-day task for multi-weather corruption restoration.
    \item Weather-RainGAN:Conditional GANs trained on rain-to-clear task for multi-weather corruption restoration.
    \item Restormer (Gaussian color denoising)  for multi-weather corruption restoration.
    \item Restormer (Gaussian grayscale denoising)  for multi-weather corruption restoration.
\end{enumerate}

\subsection{Conditional Generative Adversarial Networks}

GANs or Generative Adversarial networks are generative models that learn mapping between noisy z and output image y, G : z → y. Conditional GANs learn a mapping from observed image x and random noise vector z, to y, G : {x, z} → y \cite{a1}. The generator G is trained to produce images similar to the “real” images, as compared by an adversarially trained discriminator, D, which is used for detecting the “fakes”. The final objective of the conditional GAN can be expressed as: 

\begin{equation*}
   G^{*} = arg  min_{G} max_{D} \mathcal{L}_{cGAN} (G, D) +  \lambda \mathcal{L}_{L1} (G) 
\end{equation*}

where G tries to minimize this objective against an adversarial D that tries to maximize it, i.e. $G^{*} = arg  min_{G} max_{D} L_{cGAN} (G, D).$

In this study, we propose 2 new use-cases of the conditional GAN for restoration purposes. In the first case, the GAN is trained on night-to-day images and tested for multi-weather corruption tasks. The intuition behind this is the similarity in corruptions of night and bad weather images, like poor lighting and condensation on lens. In the second use-case, the GAN is trained on only rain-to-clear images (synthetically generated) and tested for multi-weather corruption tasks. The intuition behind this is the similarity in corruptions of rain and bad weather images, like snow and rain streaks. There is an additional challenge which notes if the single-weather trained conditional GAN can adapt to multi-weather corruptions.

\subsection{Restormer}

The restormer is a highly efficient transformer that was proposed for denoising tasks in image restoration \cite{a2}. It consists of a multi-Dconv head transposed attention (MDTA) and a gated-Dconv feed-forward network (GDFN). These proposed architectural changes gave it the ability to capture long-range pixel interactions, while still remaining applicable to large images. It is both computationally efficient, and has the capacity to handle high-resolution images, a feature critical for a task like adverse weather object detection. In this work, we study the effects of trained noise levels (15, 25 and 50) on the denoising performance of color and grayscale Restormers on the DAWN dataset. The goal is to study how the noise level affects the model performance in OD task and which level is optimal for generating explainable detections.

\subsection{Grad-CAM}

Gradient-weighted Class Activation Mapping (Grad-CAM) is a technique that produces visual explanations for the purpose of making CNN-based models more transparent \cite{a3}. For getting the class discriminative localization map Grad-CAM   ${L_{Grad-CAM}^c}$   $\epsilon  R ^ {u * v}$ of width u and height v for any class c , first comes the computation of gradient of the score for class c, $y^c$ (before the softmax), according to feature maps $A^k$ of a convolutional layer, i.e. $\frac{\partial{\mathbf{y^{c}}}}{\partial A_{}^k}$ . The neuron importance weights $a_c^k$ are attained by global-average-pooling these gradients flowing back. The ‘importance’ of feature map k for a target class c is captured by this weight $a_c^k$.  A weighted combination of forward activation maps is performed, and followed by a ReLU to obtain,
This results in a coarse heat-map of the same size as the convolutional feature maps. Grad-CAM is used for the purpose of explaining object detection in the restored images of different techniques compared in this study. We additionally use the Grad-CAM model’s detection probability in calculating the contrastive metric for monitoring training progress.

\subsection{Proposed Approach: RestoreX-AI}

Due to the instability of training of GANs, over-or-under training does not always lead to the perfect solution images for object detection. Parallelly, tuning on high noise levels does not always provide the best images from the Restormer model. Even after producing good images by standard metrics (PSNR, SSIM), their applicability for the OD task remains uncertain, which brings the need for a new evaluation standard. We propose using a weighted sum of the explainability results (detection probability of class provided by the Grad-CAM model) and the similarity of the predicted and actual label to define this new standard. This weighted sum is calculated for every stage of training (one stage can be a user-defined set of epochs), and then used to monitor the progress of the model. We introduce this new parameter for assessing the quality of restoration which is calculated using equation 1. 

\begin{equation}
    \Delta \phi  = \Delta ( \Sigma (S(p,a)*d)/N )
 \end{equation}
 
 Here S is the similarity of labels that be measured either by grouping the objects (cars, race cars and taxis have similarity 1, person, groom have 1 and so on), or by strict parameters (cars and race cars have similarity 0). p and a are the predicted and actual labels of the object under detection. N is the number of training samples generated in that stage, which are used to evaluate the current progress of that model. Refer to Appendix \ref{appendix:a} Table \ref{tab2} to view the measure of similarity of objects grouped together for this study. The explanation probability is d or the value returned by the Grad-CAM model, which is its prediction of what is present in the image. The quality of restoration between stages can be denoted as $ \Delta  \phi  $, or the difference between qualities at consecutive stages.

\subsection{Datasets}

For training the Weather-RainGAN and Weather-NightGAN, corresponding images of the same scene in rain-clear and night-day conditions were required. For the Weather-RainGAN, we used Rain 100L \cite{a32}, which is a synthesized data of rain streaks with corresponding rain-free images. For the Weather-NightGAN we used Transient Attributes dataset \cite{a34}, which used a high-level image editing method which allows a user to adjust the attributes of a scene, e.g. change a scene to be “night” or “day”. The final testing of all restoration methods required a multi-weather dataset with high-resolution images, for which the DAWN dataset was selected. The DAWN dataset is a large vehicle detection dataset which has captured images of driving scenes in adverse weather conditions \cite{a35}. It consists of 1000 images from real-traffic scenes as seen in multiple adverse weather conditions including fog, snow, rain, and sandstorms. The images have been annotated with 2D annotations(boxes) with 6 object classes namely  car, bus, truck, motorcycle, person and bicycle.

\begin{figure*}
\begin{center}
\includegraphics[width=\textwidth,height=8.5in]{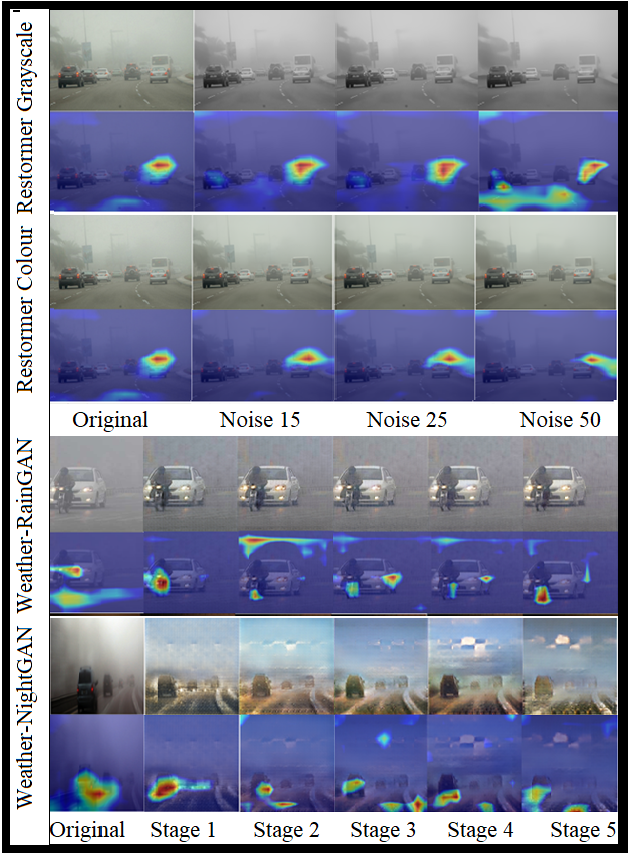}
\end{center}
  \caption{Foggy weather condition: Grad-CAM Attention Maps for (a) Restormer Grayscale Denoising (b) Restormer Colour Denoising (c) Weather-RainGAN and (d) Weather-NightGAN.}
\label{fig2}
\end{figure*}

\begin{figure*}
\begin{center}
\includegraphics[width=\textwidth,height=8.5in]{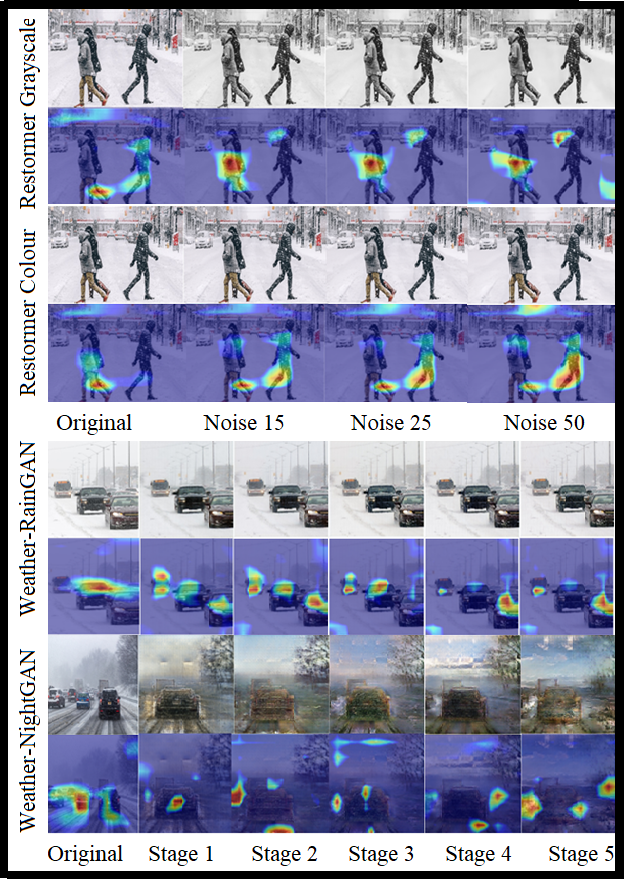}
\end{center}
  \caption{Snowfall weather condition: Grad-CAM Attention Maps for (a) Restormer Grayscale Denoising (b) Restormer Colour Denoising (c) Weather-RainGAN and (d) Weather-NightGAN.}
\label{fig3}
\end{figure*}

\begin{table*}[h]
 \centering
\begin{tabular}{llllll} \hline 
\textbf{Image Restoration Technique} & \textbf{\begin{tabular}[c]{@{}l@{}}Class 1 AP\\ {[}car{]}\end{tabular}} & \textbf{\begin{tabular}[c]{@{}l@{}}Class 2 AP\\ {[}bus{]}\end{tabular}} & \textbf{\begin{tabular}[c]{@{}l@{}}Class 3 AP\\ {[}person{]}\end{tabular}} & \textbf{\begin{tabular}[c]{@{}l@{}}Class 4 AP\\ {[}motorcycle{]}\end{tabular}} & \textbf{mAP} \\ \hline 
\multicolumn{6}{l}{\textbf{Gaussian Gray Denoising Restormer}}                                                                                                                                                                                                                                                                                                         \\\hline 
Noise 15                             & 1                                                                       & 5                                                                       & 22                                                                         & 0                                                                              & 6            \\
Noise 25                             & 1                                                                       & 9                                                                       & 23                                                                         & 0                                                                              & 6            \\
Noise 50                             & 1                                                                       & 0                                                                       & 29                                                                         & 0                                                                              & 6            \\ \hline 
\multicolumn{6}{l}{\textbf{Gaussian Color Denoising Restormer}}                                                                                                                                                                                                                                                                                                       \\ \hline  
Noise 15                             & 1                                                                       & 3                                                                       & 21                                                                         & 0                                                                              & 5            \\
Noise 25                             & 1                                                                       & 5                                                                       & 22                                                                         & 0                                                                              & 6            \\
Noise 50                             & 1                                                                       & 11                                                                      & 24                                                                         & 0                                                                              & 7            \\ \hline 
\multicolumn{6}{l}{\textbf{Weather-RainGAN}}                                                                                                                                                                                                                                                                                                                        \\\hline   
Stage 1                              & 1                                                                       & 0                                                                       & 0                                                                          & 0                                                                              & 0            \\
Stage 2                              & 1                                                                       & 17                                                                      & 0                                                                          & 0                                                                              & 4            \\
Stage 3                              & 1                                                                       & 0                                                                       & 0                                                                          & 0                                                                              & 0            \\
Stage 4                              & 1                                                                       & 0                                                                       & 0                                                                          & 76                                                                             & 15           \\
Stage 5                              & 1                                                                       & 0                                                                       & 0                                                                          & 0                                                                              & 0            \\ \hline 
\multicolumn{6}{l}{\textbf{Weather-NightGAN}}                                                                                                                                                                                                                                                                                                                          \\\hline
Stage 1                              & 11                                                                      & 0                                                                       & 0                                                                          & 0                                                                              & 2            \\
Stage 2                              & 1                                                                       & 0                                                                       & 0                                                                          & 0                                                                              & 0            \\
Stage 3                              & 1                                                                       & 0                                                                       & 0                                                                          & 0                                                                              & 0            \\
Stage 4                              & 17                                                                      & 0                                                                       & 0                                                                          & 0                                                                              & 3            \\
Stage 5                              & 48                                                                      & 0                                                                       & 0                                                                          & 0                                                                              & 10      \\ \hline    
\end{tabular}
\caption{Object detection performance of all 4 restoration methods as measured at the designed stages.}
\label{tab1}
\end{table*}

\section{Experiments and Results}

\subsection{Restormer-based Grayscale Denoising}

Section 1 of Table \ref{tab1} shows the object detection scores of Gaussian grayscale image denoising using the Restormer model. To compare the clarity and detection scores,  models trained on different noise levels 15, 25 and 50 are included in testing. The purpose of this experimentation is to identify which noise level is optimal for the object detection task, additionally verified by the explainability setup shown in Figures \ref{fig2} and \ref{fig3}. The overall mAP of object detection remains constant on all noise levels, however individual class scores are observed to fluctuate. The AP for bus and person class increases from noise 15 to 25 by 4 AP and 1AP respectively. However the bus AP drops to 0 and person AP boosts to 29 when noise is set to 50. The increase in noise level would correspond to smoother images produced by restormer, however,  it is observed that the bus AP drops when noise level is set to 50. This is very interesting to note and also observable in Figure \ref{fig2}, where the attention maps shift from the bus to the car as the noise level increases.

\subsection{Restormer-based Colour Denoising}

Section 2 of Table \ref{tab1} shows the object detection scores of Gaussian color image denoising using the Restormer model. To compare the clarity and detection scores, models trained on different noise levels 15, 25 and 50 are included in testing. The purpose of this experimentation is to identify which noise level is optimal for the object detection task, additionally verified by the explainability setup. The overall mAP of object detection increases steadily by 1 with all increasing noise levels, and individual class scores are observed to increase as well. The AP for bus and person class increases from noise 15 to 25 by 2 AP and 1AP respectively. The bus AP further increases to 11 and person AP boosts to 24 when noise is set to 50. The increase in noise level is improving the image quality and object detection. As visible in Figure \ref{fig3}, the attention maps on the people are focusing in the region of interest as the noise level increases.

\subsection{Weather-RainGAN}

The utilization of Pix2Pix GAN \cite{a1} for deraining purposes in mapped rain-clear images is proposed, and validated using our proposed technique. The dataset used for this purpose was Rain 100L \cite{a32}, which is a synthesized data of rain streaks with corresponding rain-free images. The images in Rain 100 L are originally from BSD 200 dataset \cite{a33}. The GAN model is trained with the rainy images as source and clear images as target, expecting this technique to produce denoised images on test images of the DAWN dataset. The intuition behind this idea was the inherent similarity between weather corruptions and synthetic rain streaks, with the goal of the GAN learning to work similarly on the 2 tasks. The training epochs are divided into stages(1,2,3,4,5) to monitor the training progress of the GANs and produce results of OD and explainability at each stage. As expected, the GANs are producing highly unstable behavior with increasing and 0 AP at most of the stages. However, the sudden boost in AP at specific epochs inspired the formulation of our proposed weighted explainability measure. It can be seen in Section 3 of Table \ref{tab1}, that the bus AP is 17 at Stage 4, but 0 at all other stages and the motorcycle AP is 76 at Stage 8 and 0 at other Stages. The car AP however, remains constant at 1 and overall mAP increases over time due to the individual class performance boosts. As these results are vague, we take a closer look at Figures \ref{fig2} and \ref{fig3}, to determine possible causes for this object detection performance. It can be observed that  the GAN is actually performing quite well in denoising weather conditions like fog and snow, and the attention maps are converging towards the relevant objects as the training progresses. But while the object detection results are very disjoint, the attention maps progress continuously and show improvement. The effectiveness of this GAN can be observed as a deraining solution as the image not only gets clearer, but the object detection and attention maps get more precise over training as well.

\subsection{Weather-NightGAN}

A new use-case of the PixtoPix GAN \cite{a1} is proposed, trained on a different use-case for deraining purposes in mapped rain-clear images, and validated our proposed technique using the experimental procedure. The dataset used for this purpose was  Transient Attributes dataset \cite{a34}, which used a high-level image editing method which allows a user to adjust the attributes of a scene, e.g. change a scene to be “night” or “day”. The GAN model is trained with the night images as source and day images as target, expecting this technique to produce denoised images on test images of the DAWN dataset. The intuition behind this idea was the inherent similarity between weather corruptions and dark night images. As expected, the GANs are producing highly unstable behavior with increasing and 0 AP at most of the stages as shown in Table \ref{tab1}. This technique actually worked out only for 1 class (car) and boosted its AP from 11 to 48 over the training period. The remaining classes had a constant 0 AP. By taking a closer look at Figures \ref{fig2} and \ref{fig3}, possible causes for this object detection performance can be determined. The GAN is actually performing quite randomly in denoising weather conditions like fog and snow using night training images, but the attention maps are still converging towards the relevant objects as the training progresses. This is exactly aligned with the initial analysis and intuition for the discovery i.e. cases in which object detection is clear to a CV detector, although imperceptible to the human eye. While the object detection results are very disjoint, the attention maps progress continuously and show improvement. The effectiveness of this GAN can be observed in denoising weather conditions for cars, but not for other classes. The purpose of our methodology is not just to observe the restoration progress in a positive light, but to also stop training in case the GAN goes too far. As can be seen in this particular case, the GAN distorts the images after Stage 2, which may confuse detectors when tested against it. In Stage 4, the GAN produces images which are heavily distorted but oddly easier for car detection than even the original image. 

The attention maps and results for all four restoration techniques and four weather conditions: fog, rain, snow and dust tornado have been displayed in Appendix \ref{appendix:a}.

\section{Conclusion}

The goal of our work was to study the relationship between object detection performance, image clarity, explainability and training time in the context of novel versus established restoration techniques. The scope of this study covered 4 different restoration techniques, aiming to denoise images corrupted by over 6 different weather conditions presented in the DAWN dataset, namely fog, snow storm, haze, dust tornadoes, rainfall and mist. All 4 techniques worked differently, with the Restormer-based methods providing stable all-rounded results, while the GANs provided class-specific boosts in performance. The overall rise in mAP before and after applying the techniques was 0\%, 40\%, 275\%, 400\% respectively. Contrary to popular beliefs, greater denoising does not always guarantee better object detection results as observed in this paper. And conversely, poor denoising does not always guarantee worse object detection results. Particularly for specific tasks like bus and car detection, it can be seen that newer approaches like our proposed method (Weather-RainGAN and Weather-NightGAN) can boost the detector’s performance with its resultant images. We present conditional GANs cases that perform superbly on diverse weather conditions ranging from dust tornadoes to snowfall, in spite of having trained on limited single weather conditions.

We present 2 very interesting observations obtained through this study:
\begin{enumerate}
    \item GANs can generate images complex to the human eye, but comparatively interpretable for vision models post processing. This opens the possibility of exploring the capabilities of vision models beyond the scope of human vision and also warding off potential attacks which can cripple modern detectors. 
    \item Restoration and denoising methods which produce clearer images (as measured using standard image quality metrics like PSNR), may in fact present a greater challenge for machine perception in object detection.
    
\end{enumerate}
Understanding how differently humans and detectors perceive information in the same image demands greater exploration of explainability. We would like to open the discussion for countless possibilities arising from these disparate perspectives, E.g. if a model sees a pedestrian on a rainy road which a human cannot see, or conversely  a pedestrian visible to a passenger’s eye which a model cannot capture. While using denoising techniques are a popular choice for tackling corruptions, it must also be acknowledged that not all techniques are suitable for all use-cases as seen for car detection. Certain classes may respond better to denoising depending on their physical characteristics as perceived by the detectors. Going ahead with building robust, explainable models, this problem must be studied from multiple perspectives in the future. We hope to inspire a new line of research in this direction which deals with the complexity of the task of weather corruptions and can solve it using diverse generalizable solutions.

{\small
\bibliographystyle{ieee_fullname}
\bibliography{egbib}

\begin{thebibliography}{10}\itemsep=-1pt

\bibitem{a15}
Sebastian Bach, Alexander Binder, Gr{\'e}goire Montavon, Frederick Klauschen,
  Klaus-Robert M{\"u}ller, and Wojciech Samek.
\newblock On pixel-wise explanations for non-linear classifier decisions by
  layer-wise relevance propagation.
\newblock {\em PloS one}, 10(7):e0130140, 2015.

\bibitem{a16}
David Baehrens, Timon Schroeter, Stefan Harmeling, Motoaki Kawanabe, Katja
  Hansen, and Klaus-Robert M{\"u}ller.
\newblock How to explain individual classification decisions.
\newblock {\em The Journal of Machine Learning Research}, 11:1803--1831, 2010.

\bibitem{a6}
Mario Bijelic, Tobias Gruber, Fahim Mannan, Florian Kraus, Werner Ritter, Klaus
  Dietmayer, and Felix Heide.
\newblock Seeing through fog without seeing fog: Deep multimodal sensor fusion
  in unseen adverse weather.
\newblock In {\em Proceedings of the IEEE/CVF Conference on Computer Vision and
  Pattern Recognition}, pages 11682--11692, 2020.

\bibitem{a5}
Mario Bijelic, Tobias Gruber, and Werner Ritter.
\newblock Benchmarking image sensors under adverse weather conditions for
  autonomous driving.
\newblock In {\em 2018 IEEE Intelligent Vehicles Symposium (IV)}, pages
  1773--1779. IEEE, 2018.

\bibitem{a11}
Antoni Buades, Bartomeu Coll, and Jean-Michel Morel.
\newblock A review of image denoising algorithms, with a new one.
\newblock {\em Multiscale modeling \& simulation}, 4(2):490--530, 2005.

\bibitem{a4}
Pak~Hung Chan, Gunwant Dhadyalla, and Valentina Donzella.
\newblock A framework to analyze noise factors of automotive perception
  sensors.
\newblock {\em IEEE Sensors Letters}, 4(6):1--4, 2020.

\bibitem{a27}
Ke Chen, Ryan Oldja, Nikolai Smolyanskiy, Stan Birchfield, Alexander Popov,
  David Wehr, Ibrahim Eden, and Joachim Pehserl.
\newblock Mvlidarnet: Real-time multi-class scene understanding for autonomous
  driving using multiple views.
\newblock In {\em 2020 IEEE/RSJ International Conference on Intelligent Robots
  and Systems (IROS)}, pages 2288--2294. IEEE, 2020.

\bibitem{a37}
Alain Hore and Djemel Ziou.
\newblock Image quality metrics: Psnr vs. ssim.
\newblock In {\em 2010 20th international conference on pattern recognition},
  pages 2366--2369. IEEE, 2010.

\bibitem{a1}
Phillip Isola, Jun-Yan Zhu, Tinghui Zhou, and Alexei~A Efros.
\newblock Image-to-image translation with conditional adversarial networks.
\newblock In {\em Proceedings of the IEEE conference on computer vision and
  pattern recognition}, pages 1125--1134, 2017.

\bibitem{a35}
Mourad~A Kenk and Mahmoud Hassaballah.
\newblock Dawn: vehicle detection in adverse weather nature dataset.
\newblock {\em arXiv preprint arXiv:2008.05402}, 2020.

\bibitem{a19}
Rajiv Khanna, Been Kim, Joydeep Ghosh, and Sanmi Koyejo.
\newblock Interpreting black box predictions using fisher kernels.
\newblock In {\em The 22nd International Conference on Artificial Intelligence
  and Statistics}, pages 3382--3390. PMLR, 2019.

\bibitem{a20}
Pang~Wei Koh and Percy Liang.
\newblock Understanding black-box predictions via influence functions.
\newblock In {\em International conference on machine learning}, pages
  1885--1894. PMLR, 2017.

\bibitem{a26}
Jason Ku, Melissa Mozifian, Jungwook Lee, Ali Harakeh, and Steven~L Waslander.
\newblock Joint 3d proposal generation and object detection from view
  aggregation.
\newblock In {\em 2018 IEEE/RSJ International Conference on Intelligent Robots
  and Systems (IROS)}, pages 1--8. IEEE, 2018.

\bibitem{a34}
Pierre-Yves Laffont, Zhile Ren, Xiaofeng Tao, Chao Qian, and James Hays.
\newblock Transient attributes for high-level understanding and editing of
  outdoor scenes.
\newblock {\em ACM Transactions on Graphics (proceedings of SIGGRAPH)}, 33(4),
  2014.

\bibitem{a25}
Peiliang Li, Xiaozhi Chen, and Shaojie Shen.
\newblock Stereo r-cnn based 3d object detection for autonomous driving.
\newblock In {\em Proceedings of the IEEE/CVF Conference on Computer Vision and
  Pattern Recognition (CVPR)}, June 2019.

\bibitem{a22}
Wei Liu, Dragomir Anguelov, Dumitru Erhan, Christian Szegedy, Scott Reed,
  Cheng-Yang Fu, and Alexander~C Berg.
\newblock Ssd: Single shot multibox detector.
\newblock In {\em European conference on computer vision}, pages 21--37.
  Springer, 2016.

\bibitem{a30}
Aboli Marathe, Rahee Walambe, and Ketan Kotecha.
\newblock Evaluating the performance of ensemble methods and voting strategies
  for dense 2d pedestrian detection in the wild.
\newblock In {\em Proceedings of the IEEE/CVF International Conference on
  Computer Vision}, pages 3575--3584, 2021.

\bibitem{a33}
David Martin, Charless Fowlkes, Doron Tal, and Jitendra Malik.
\newblock A database of human segmented natural images and its application to
  evaluating segmentation algorithms and measuring ecological statistics.
\newblock In {\em Proceedings Eighth IEEE International Conference on Computer
  Vision. ICCV 2001}, volume~2, pages 416--423. IEEE, 2001.

\bibitem{a18}
Gr{\'e}goire Montavon, Wojciech Samek, and Klaus-Robert M{\"u}ller.
\newblock Methods for interpreting and understanding deep neural networks.
\newblock {\em Digital Signal Processing}, 73:1--15, 2018.

\bibitem{a36}
Valentina Mușat, Ivan Fursa, Paul Newman, Fabio Cuzzolin, and Andrew Bradley.
\newblock Multi-weather city: Adverse weather stacking for autonomous driving.
\newblock In {\em Proceedings of the IEEE/CVF International Conference on
  Computer Vision}, pages 2906--2915, 2021.

\bibitem{a17}
Anh Nguyen, Alexey Dosovitskiy, Jason Yosinski, Thomas Brox, and Jeff Clune.
\newblock Synthesizing the preferred inputs for neurons in neural networks via
  deep generator networks.
\newblock {\em Advances in neural information processing systems}, 29, 2016.

\bibitem{a7}
Vladislav Ostankovich, Rauf Yagfarov, Maksim Rassabin, and Salimzhan Gafurov.
\newblock Application of cyclegan-based augmentation for autonomous driving at
  night.
\newblock In {\em 2020 International Conference Nonlinearity, Information and
  Robotics (NIR)}, pages 1--5. IEEE, 2020.

\bibitem{a9}
Horia Porav, Tom Bruls, and Paul Newman.
\newblock I can see clearly now: Image restoration via de-raining.
\newblock In {\em 2019 International Conference on Robotics and Automation
  (ICRA)}, pages 7087--7093. IEEE, 2019.

\bibitem{a8}
Horia Porav, Will Maddern, and Paul Newman.
\newblock Adversarial training for adverse conditions: Robust metric
  localisation using appearance transfer.
\newblock In {\em 2018 IEEE international conference on robotics and automation
  (ICRA)}, pages 1011--1018. IEEE, 2018.

\bibitem{a24}
Joseph Redmon, Santosh Divvala, Ross Girshick, and Ali Farhadi.
\newblock You only look once: Unified, real-time object detection.
\newblock In {\em Proceedings of the IEEE conference on computer vision and
  pattern recognition}, pages 779--788, 2016.

\bibitem{a23}
Shaoqing Ren, Kaiming He, Ross Girshick, and Jian Sun.
\newblock Faster r-cnn: Towards real-time object detection with region proposal
  networks.
\newblock {\em Advances in neural information processing systems}, 28, 2015.

\bibitem{a21}
Ramprasaath~R Selvaraju, Michael Cogswell, Abhishek Das, Ramakrishna Vedantam,
  Devi Parikh, and Dhruv Batra.
\newblock Grad-cam: Visual explanations from deep networks via gradient-based
  localization.
\newblock In {\em Proceedings of the IEEE international conference on computer
  vision}, pages 618--626, 2017.

\bibitem{a3}
Ramprasaath~R Selvaraju, Michael Cogswell, Abhishek Das, Ramakrishna Vedantam,
  Devi Parikh, and Dhruv Batra.
\newblock Grad-cam: Visual explanations from deep networks via gradient-based
  localization.
\newblock In {\em Proceedings of the IEEE international conference on computer
  vision}, pages 618--626, 2017.

\bibitem{a14}
Avanti Shrikumar, Peyton Greenside, Anna Shcherbina, and Anshul Kundaje.
\newblock Not just a black box: Learning important features through propagating
  activation differences.
\newblock {\em arXiv preprint arXiv:1605.01713}, 2016.

\bibitem{a12}
Karen Simonyan, Andrea Vedaldi, and Andrew Zisserman.
\newblock Deep inside convolutional networks: Visualising image classification
  models and saliency maps.
\newblock {\em arXiv preprint arXiv:1312.6034}, 2013.

\bibitem{a29}
Rahee Walambe, Aboli Marathe, and Ketan Kotecha.
\newblock Multiscale object detection from drone imagery using ensemble
  transfer learning.
\newblock {\em Drones}, 5(3):66, 2021.

\bibitem{a31}
Rahee Walambe, Aboli Marathe, Ketan Kotecha, and George Ghinea.
\newblock Lightweight object detection ensemble framework for autonomous
  vehicles in challenging weather conditions.
\newblock {\em Computational Intelligence and Neuroscience}, 2021, 2021.

\bibitem{a32}
Wenhan Yang, Robby~T Tan, Jiashi Feng, Jiaying Liu, Zongming Guo, and Shuicheng
  Yan.
\newblock Deep joint rain detection and removal from a single image.
\newblock In {\em Proceedings of the IEEE conference on computer vision and
  pattern recognition}, pages 1357--1366, 2017.

\bibitem{a10}
Syed~Waqas Zamir, Aditya Arora, Salman Khan, Munawar Hayat, Fahad~Shahbaz Khan,
  and Ming-Hsuan Yang.
\newblock Restormer: Efficient transformer for high-resolution image
  restoration.
\newblock {\em arXiv preprint arXiv:2111.09881}, 2021.

\bibitem{a2}
Syed~Waqas Zamir, Aditya Arora, Salman Khan, Munawar Hayat, Fahad~Shahbaz Khan,
  and Ming-Hsuan Yang.
\newblock Restormer: Efficient transformer for high-resolution image
  restoration.
\newblock {\em arXiv preprint arXiv:2111.09881}, 2021.

\bibitem{a13}
Matthew~D Zeiler and Rob Fergus.
\newblock Visualizing and understanding convolutional networks.
\newblock In {\em European conference on computer vision}, pages 818--833.
  Springer, 2014.

\bibitem{a28}
Yin Zhou, Pei Sun, Yu Zhang, Dragomir Anguelov, Jiyang Gao, Tom Ouyang, James
  Guo, Jiquan Ngiam, and Vijay Vasudevan.
\newblock End-to-end multi-view fusion for 3d object detection in lidar point
  clouds.
\newblock In {\em Conference on Robot Learning}, pages 923--932. PMLR, 2020.

\end{thebibliography}
}

\newpage

\appendix

\section{Appendix}
\label{appendix:a}

The table \ref{tab2} contains the categories of labels grouped with a similarity measure of 1 in the proposed approach.

\begin{table*}[h]
 \centering
\begin{tabular}{lllll}
\hline
\textbf{Person} & \textbf{Car}  & \textbf{Motorcycle} & \textbf{Bus} & \textbf{Bicycle} \\ \hline
Groom           & Cab           & Moped               & Trolley bus  & Tandem bicycle   \\
Bridegroom      & Taxi          &                     & Mini bus     & Tricycle         \\
Baseball player & Race car      &                     & School bus   & Unicycle         \\
Scuba Diver     & Jeep          &                     &              & Mountain bike    \\
                & Minivan       &                     &              & All-terrain bike \\
                & Estate car    &                     &              & Off-roader       \\
                & Station wagon &                     &              & Trike            \\ \hline
\end{tabular}
\caption{Categories of labels grouped with Similarity 1.}

\label{tab2}
\end{table*}

Given below are the extra figures to supplement the experimentation section of the paper.

\begin{figure*}
\begin{center}
\includegraphics[width=\textwidth,height=8.5in]{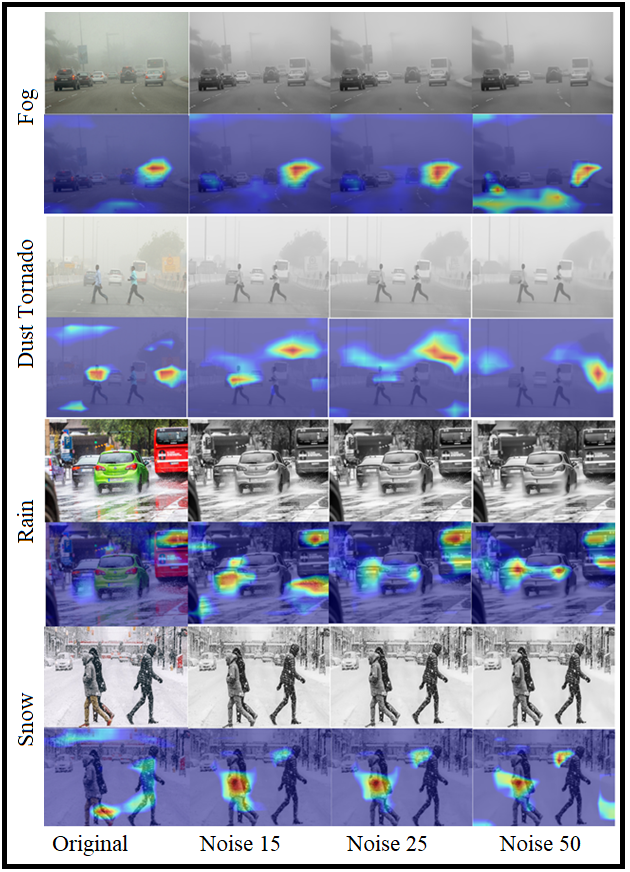}
\end{center}
  \caption{Restormer Grayscale Denoising: Grad-CAM Attention Maps for (a) Fog (b) Dust tornado (c) Rain and (d) Snow weather conditions.}
\label{fig4}
\end{figure*}

\begin{figure*}[h]
\begin{center}
\includegraphics[width=\textwidth,height=8.5in]{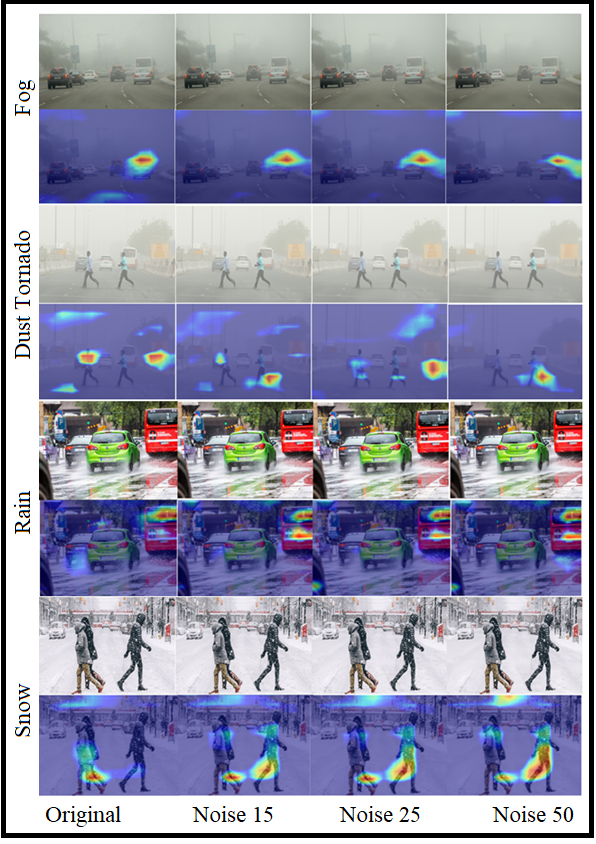}
\end{center}
  \caption{Restormer Color Denoising: Grad-CAM Attention Maps for (a) Fog (b) Dust tornado (c) Rain and (d) Snow weather conditions.}
\label{fig5}
\end{figure*}

\begin{figure*}
\begin{center}
\includegraphics[width=\textwidth,height=8.5in]{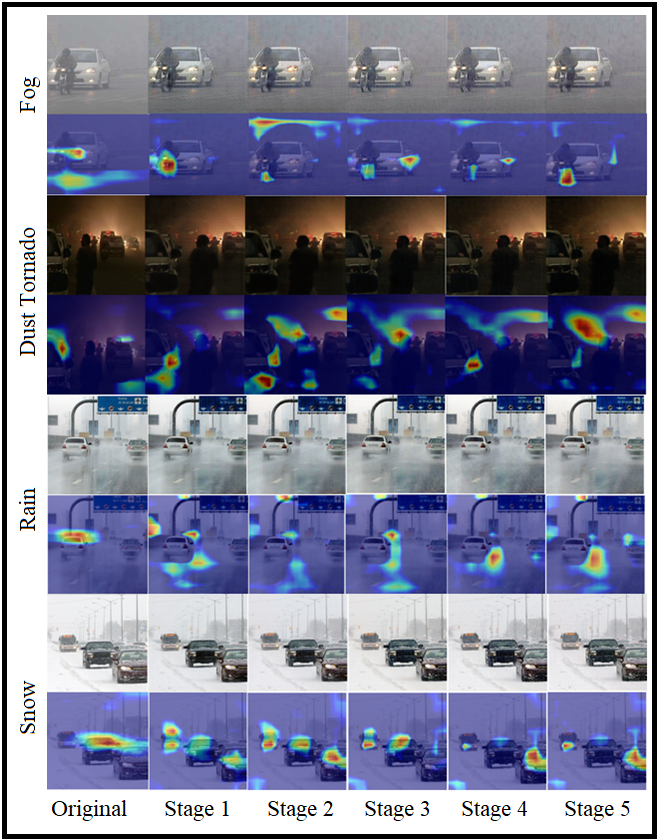}
\end{center}
  \caption{Weather-RainGAN: Grad-CAM Attention Maps for (a) Fog (b) Dust tornado (c) Rain and (d) Snow weather conditions.}
\label{fig6}
\end{figure*}

\begin{figure*}[h]
\begin{center}
\includegraphics[width=\textwidth,height=8.5in]{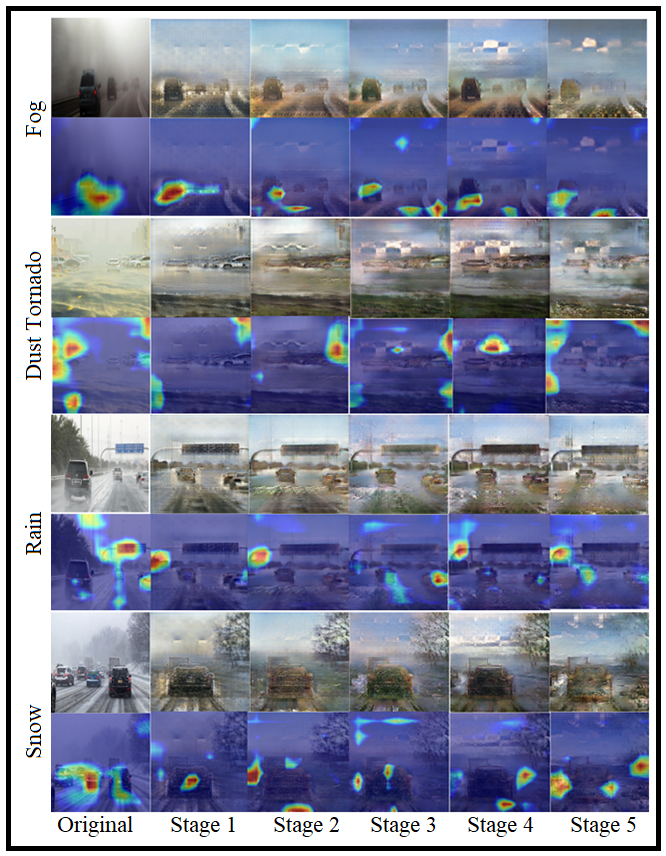}
\end{center}
  \caption{Weather-NightGAN: Grad-CAM Attention Maps for (a) Fog (b) Dust tornado (c) Rain and (d) Snow weather conditions.}
\label{fig7}
\end{figure*}

\end{document}